\documentclass[acmsmall,screen]{acmart}

\AtBeginDocument{
  \providecommand\BibTeX{{\normalfont B\kern-0.5em{\scshape i\kern-0.25em b}\kern-0.8em\TeX}}
}

\makeatletter
\def\@ACM@checkaffil{
    \if@ACM@instpresent\else
    \ClassWarningNoLine{\@classname}{No institution present for an affiliation}%
    \fi
    \if@ACM@citypresent\else
    \ClassWarningNoLine{\@classname}{No city present for an affiliation}%
    \fi
    \if@ACM@countrypresent\else
        \ClassWarningNoLine{\@classname}{No country present for an affiliation}%
    \fi
}
\makeatother

\usepackage{microtype} 
\usepackage{hyperref}
\usepackage{cite}
\usepackage{amsmath,amsfonts}
\usepackage{graphicx}
\usepackage{textcomp}
\usepackage{xcolor}
\usepackage{comment}
\usepackage[ruled,vlined,linesnumbered]{algorithm2e}
\usepackage{setspace}
\usepackage{xspace}
\usepackage{enumitem}
\usepackage{url}
\usepackage{multirow}
\usepackage{hyperref}
\usepackage{booktabs}
\usepackage{tcolorbox}
\usepackage{bbm}
\usepackage{balance}
\usepackage{tabularx}
\usepackage{diagbox}
\usepackage{listings}
\usepackage[utf8]{inputenc}
\usepackage[T1]{fontenc}
\usepackage{subcaption}

\usepackage{multicol}
\usepackage{makecell}
\usepackage{pifont}
\usepackage{wrapfig}

\DeclareUnicodeCharacter{2208}{\ensuremath{\in}}
\usepackage[ruled,vlined]{algorithm2e}

\usepackage[frozencache,cachedir=.]{minted}
\setminted{
    breaklines=true, 
    baselinestretch=1.0, 
    fontsize=\footnotesize, 
    linenos, 
    numbersep=5pt, 
    tabsize=4, 
    obeytabs, 
}

\newminted{java}{
    style=default,
    bgcolor=lightgray,
}

\lstdefinestyle{Java}{
    language=Java, 
    basicstyle=\ttfamily\small, 
    keywordstyle=\color{blue}, 
    commentstyle=\color{green!50!black}, 
    stringstyle=\color{red}, 
    showstringspaces=false, 
    breaklines=true, 
    frame=single
}

\newcommand{\toolname}[0]{\textsc{VulDebugger}\xspace}

\newcommand{\ExtractFix}[0]{\textsc{ExtractFix}\xspace}
\newcommand{\Vrepair}{\textsc{Vrepair}\xspace}
\newcommand{\VulRepair}{\textsc{VulRepair}\xspace}

\newcommand{\autocoderover}[0]{\textsc{AutoCodeRover}\xspace}

\newcommand{\MarsCodeAgent}[0]{\textsc{MarsCode Agent}\xspace}

\newcommand{\projectprecision}[0]{60.00\%\xspace}

\newcommand{\tick}[0]{\ding{51}}
\newcommand{\cross}[0]{\ding{55}}

\theoremstyle{definition}

\begin{document}

\title{Agent That Debugs: Dynamic State-Guided Vulnerability Repair}


\author{Zhengyao Liu}
\affiliation{
  \institution{Beihang University}
}
\email{zhengyaoliu@buaa.edu.cn}

\author{Yunlong Ma}
\affiliation{
  \institution{Beihang University}
}
\email{yunlong_ma@buaa.edu.cn}

\author{Jingxuan Xu}
\affiliation{
  \institution{Beihang University}
}
\email{jingxuan_xu@buaa.edu.cn}

\author{Junchen Ai}
\affiliation{
  \institution{Beihang University}
}
\email{junchen_ai@buaa.edu.cn}

\author{Xiang Gao}
\affiliation{
  \institution{Beihang University}
}
\email{xiang_gao@buaa.edu.cn}

\author{Hailong Sun}
\affiliation{
  \institution{Beihang University}
}
\email{sunhl@buaa.edu.cn}

\author{Abhik Roychoudhury}
\affiliation{
  \institution{National University of Singapore}
}
\email{abhik@comp.nus.edu.sg}

\begin{abstract}

In recent years, more vulnerabilities have been discovered every day, while manual vulnerability repair requires specialized knowledge and is time-consuming.
As a result, many detected or even published vulnerabilities remain unpatched, thereby increasing the exposure of software systems to attacks.
Recent advancements in agents based on Large Language Models have demonstrated their increasing capabilities in code understanding and generation, which can be promising to achieve automated vulnerability repair.
However, the effectiveness of agents based on static information retrieval is still not sufficient for patch generation.
To address the challenge, we propose a program repair agent called \toolname that fully utilizes both static and dynamic context, and it debugs programs in a manner akin to humans.
The agent inspects the actual state of the program via the debugger and infers expected states via constraints that need to be satisfied. 
By continuously comparing the actual state with the expected state, it deeply understands the root causes of the vulnerabilities and ultimately accomplishes repairs.
We experimentally evaluated \toolname on 50 real-life projects.
With \projectprecision successfully fixed, \toolname significantly outperforms state-of-the-art approaches for vulnerability repair.

\end{abstract}

\maketitle

\section{Introduction}
\label{sec:intro}

Software vulnerabilities are defects in programs that attackers can exploit to gain unauthorized access or trigger unintended behaviors~\citep{NVD}, leading to financial loss, leakage of confidential information~\citep{dowd2006art}, etc. 
According to statistics from CVE Details, the number of vulnerabilities has been steadily increasing in recent years, with 40,296 new vulnerabilities reported in 2024, and this number is expected to rise even further in 2025~\citep{CVE_Details}.
This trend indicates that software vulnerabilities have become a significant source of security threats.
However, manually fixing vulnerabilities is a challenging task that requires time and specialized expertise.
Statistics from the Edgescan report~\citep{edgescan} indicate that the average remediation time for critical-severity vulnerabilities is 65 days, leaving programs exposed to potential attacks.
Therefore, there is an urgent need for automated vulnerability repair techniques to enhance software security. 

In the past, automated vulnerability repair was a challenging task due to the diversity of vulnerability types, complexity of trigger conditions, difficulties in verification, and limitations in code generation capabilities.
Recently, with transformer-based pre-trained models~\citep{feng2020codebert, wang2021codet5, guo2022unixcoder} showing promising results in code understanding and generation, researchers have proposed several AI-based approaches like \Vrepair~\citep{chen2022neural} and \VulRepair~\citep{fu2022vulrepair} for automated vulnerability repair.
These methods leverage vulnerability datasets, such as CVEfixes~\citep{bhandari2021cvefixes} and BigVul~\citep{Fan_Li_Wang_Nguyen_2020}, to train models or fine-tune existing pre-trained models.
The goal is to obtain a model capable of understanding software vulnerabilities, which can then be used to generate repaired code or patches.
Although these methods obtain promising results, their accuracy remains relatively low, typically not exceeding 25\%.
This could be attributed to the following factors.

\begin{itemize}[leftmargin=*]
    \item \textbf{Limited Dataset}
    The effectiveness of these methods is closely tied to the quality of the dataset.
    Models trained on specific datasets often struggle to generalize to broader vulnerability repair tasks~\citep{xia2023automated}. 
    Also, the quality of existing vulnerability datasets is not optimal.
    According to VulGen~\citep{nong2023vulgen}, most high-quality vulnerability datasets are relatively small in scale.
    Moreover, existing large-scale datasets, often including significant inaccuracies and noises, fail to represent real-world vulnerabilities accurately. 
    \item \textbf{Limited Model Capability}
    Vulnerability repair is a complex task that involves multiple stages, including fault localization, root cause analysis, fix localization, and patch generation. 
    There are reasonable concerns about whether pre-trained models like CodeT5, which is used by \Vrepair~\citep{chen2022neural} and \VulRepair~\citep{fu2022vulrepair},  have sufficient parameters to handle them. 
\end{itemize} 

In recent years, Large Language Models (LLMs)~\citep{github_copilot, openai_website} have gained significant attention from researchers. 
In the field of program analysis, LLMs have also demonstrated strong capabilities in code understanding and generation, showing more promising results than traditional pre-trained models.
Consequently, a growing number of research efforts have focused on LLM-based bug detection, reasoning, debugging, and etc~\citep{fu2023chatgpt, cheshkov2023evaluation, sun2024llm4vuln, zhang2024prompt, pearce2023examining, ni2024next, kang2023explainable, zhang2024autocoderover, bouzenia2024repairagent,HOU2025107671}.
There have also been several attempts to apply LLMs to repair tasks. 
For instance, Pearce et al.~\citep{pearce2023examining} first evaluate the performance of LLMs in the zero-shot generation of security fixes, exploring the ability of LLMs to repair vulnerabilities directly.
The zero-shot approach treats LLMs as enhanced pre-trained models, but they remain limited by the training data and exhibit weak performance in repairing vulnerabilities that have not been seen before.
Moreover, LLMs are not specifically trained for vulnerability repair tasks, meaning this approach fails to fully leverage their capabilities, leaving significant room for improvement. 
The LLM4CVE framework (Fakih et al., 2024) ~\citep{Fakih2025LLM4CVEEI} creates an automated, iterative process for a Large Language Model to systematically correct vulnerabilities in code, improving on current automated vulnerability correction tools.
However, it requires manual input thus making each vulnerability repair highly resource-intensive.

Moreover, researchers~\citep{bouzenia2024repairagent, zhang2024autocoderover, HOU2025107671} propose to guide the LLM to repair defects via multiple agents.
Such approaches employ toolset~\citep{qian2023creator} to provide the LLM with various static information about source codes, treating the LLM as an interactive agent rather than merely a generation tool, thereby making more comprehensive use of its capabilities. 
However, despite static analysis providing additional information beyond the buggy code snippet, it still lacks certain critical details. 
For example, vulnerability CVE-2016-3623~\citep{cve-2016-3623}, a division-by-zero vulnerability, involves an erroneously zeroed variable \texttt{vertSubSampling} being passed through multiple function calls and several conditional checks.
This complexity makes it challenging to determine precisely where \texttt{vertSubSampling} is zeroed and how the zeroed values are propagated using static information alone. 
Similarly, issues involving buffer/heap overflows, and other problems that are traditionally difficult for static analysis to address, are also problematic to inform the LLM solely with static information. 


To address the aforementioned challenges, our key idea is to design an LLM agent that debugs programs like human developers.
Think of how we fix a vulnerability.
Given a vulnerability,
(1) we usually first read the error report and identify suspicious locations to set breakpoints.
(2) With the expected states of the program in mind, we then run the program to these locations to check the actual states.
(3) Finally, we come up with a patch to fix the discrepancies between actual and expected states.
The essence of this process involves a step-by-step analysis of the root cause of the error, focusing on continually comparing the dynamic program information with the expected state. 
This process is key to understanding and resolving the vulnerability effectively.
By guiding the LLM to compare dynamic program information with the expected state, it can also enhance its understanding of vulnerabilities through continuous comparison.
Based on this enhanced understanding, the LLM can more effectively trace the root cause of the vulnerability, identify the appropriate repair location, and generate a patch that addresses the underlying issue.

In this paper, we propose a repair agent that fully utilizes both static and dynamic context.
Similar to existing work~\citep{zhang2024autocoderover, bouzenia2024repairagent}, the static context includes the vulnerable code snippet, the information of the variable, the function body, etc.
In contrast, the dynamic context mainly involves reasoning about the actual and expected program states at certain program locations.
By setting breakpoints, we can pause the execution of the program at specific lines to obtain the program's actual state from the stack frames.
In this process, inferring the expected state is one of the biggest challenges.
To solve this problem, inspired by the crash-free constraints~\citep{gao2021beyond}, i.e., vulnerability-free constraints, we propose to infer expected states via constraints that need to be satisfied to fix the vulnerabilities.
Specifically, this approach extracts ``crash-free'' constraints at the ``crash'' location (using sanitizers to trigger a crash for a vulnerability) that can disable the observed vulnerabilities. 
These constraints are well-suited to represent the expected state associated with a vulnerability.
Furthermore, relying on the crash-free constraints, we infer the vulnerability-related states at various program locations step by step.


To realize this idea, we implement a tool called \toolname, which utilizes LLMs to debug and automatically repair vulnerabilities. 
\toolname initially triggers the vulnerability by executing a Proof of Concept (POC) to obtain fundamental crash location and crash constraint information.
Then, it directs the LLM to set breakpoints at various locations based on the crash message and source code.
At the same time, \toolname infers the constraints at these breakpoints, relays the dynamic information and expected states to the LLM, and guides it for root cause analysis and fix localization.
Finally, when the LLM gathers enough information, it tries to generate a patch and \toolname validates it by testing whether it still triggers the vulenerability or not.

The contributions of this paper are summarized as follows:

\begin{itemize}[leftmargin=*]
    \item We are the first to propose a method that enables the LLM to perform automated vulnerability repair through dynamic information analysis, introducing an innovative concept of utilizing LLMs to debug programs.
    \item We propose a repair agent that continuously compares the expected states perceived based on crash-free constraints with the actual states informed by dynamic information, facilitating the completion of automatic vulnerability repair tasks.
    \item We implement a tool called \toolname, and evaluations on real-life vulnerabilities show that \toolname outperforms existing techniques.
\end{itemize}

\clearpage
\section{Motivation}

In recent years, LLMs have demonstrated promising abilities in understanding and generating code.
Consequently, applying LLMs to vulnerability repair tasks seems to be straightforward.
However, repairing vulnerabilities using simple zero-shot methods may place excessive demands on the LLM.
\citet{fu2023chatgpt} attempts to generate patches for vulnerabilities using ChatGPT directly, which shows limited effectiveness.
The poor performance may be caused by the diverse nature and complexity of vulnerabilities, as well as the limitations of the training data in covering all types of vulnerability repairs. 
In this section, we will demonstrate our motivation by giving examples of directly using the LLM for vulnerability repair.

\subsection{Zero-shot repairs lack precision}

\begin{wrapfigure}{r}{0.61\textwidth}
\vspace{-12pt}
\begin{minted}[xleftmargin=5pt, linenos, breaklines, escapeinside=||, fontsize=\footnotesize]{C}
...
unsigned char *srcbuffs[MAX_SAMPLES];  |\label{lstline:define}|
...
- for (s = 0; s < spp; s++){ 
+ for (s = 0; s < spp && s < MAX_SAMPLES; s++){  |\label{lstline:fix}|
    /* Read each plane of a tile set into srcbuffs[s] */
    tbytes = TIFFReadTile(in, srcbuffs[s], col, row, 0, s); |\label{lstline:use}|
...
\end{minted}
\vspace{-16pt}
\captionsetup{type=listing}
\captionof{listing}{Code snippet of libtiff CVE-2016-5321}
\label{CVE-2016-5321}
\end{wrapfigure}

Listing~\ref{CVE-2016-5321} shows a code snippet from the libtiff~\citep{libtiff} project, containing a vulnerability identified as CVE-2016-5321~\citep{cve-2016-5321}. 
In this code, an array of pointers with a size of \texttt{MAX\_SAMPLES} is declared (line~\ref{lstline:define}).
However, there is no check to ensure that \texttt{s} is less than \texttt{MAX\_SAMPLES} before accessing \texttt{srcbuffs[s]} (line~\ref{lstline:use}).
This oversight leads to undefined memory access, resulting in erroneous program behavior.
The fix is straightforward, requiring an exit condition \texttt{s < MAX\_SAMPLES} in the loop (line~\ref{lstline:fix}).
This condition ensures that \texttt{s} does not exceed the declared length of the \texttt{srcbuffs} array.

Submitting the vulnerable function to GPT-4, it suggests that the line \texttt{tbuff = (unsigned char *)\_TIFFmalloc(tilesize + 8);} implies that ``\textit{Here, \texttt{tbuff} allocates \texttt{tilesize} + 8 bytes. The additional 8 bytes appear to be intended for padding or to prevent buffer overflow; however, there is no explicit justification in the code for why these 8 bytes are added, nor is there boundary checking when this buffer is utilized.}''
Nevertheless, the code shows that \texttt{tbuff} is just an intermediary for releasing pointers in \texttt{srcbuffs}, receiving values exclusively from \texttt{srcbuffs}, which in turn are initialized via \texttt{tbuff}. 
Thus, there is no buffer overflow risk, and the patches are entirely incorrect due to misjudgment.

Furthermore, we assist GPT-4 to better understand and fix the vulnerability by providing additional information. 
Table~\ref{tab:GPT-4's analysis of the vulnerability} shows how its analysis varied with different levels of information.
We utilized five commonly used types of information as aids: type of vulnerability, crash location, error message, crash constraint, and POC.
These were provided to GPT-4 in various combinations to assess the accuracy of the key outputs: analysis, repair strategy, and patch. 
Results show that individual aids rarely improve its repair ability, and even with full information, it failed to generate a correct patch.
We also evaluated a range of vulnerabilities of different types and complexities and observed similar outcomes.
For particularly obscure and complex vulnerabilities, even preliminary causal analysis has been proven to be challenging for LLMs.

\begin{table}[th]
    \caption{
    GPT-4's analysis of the vulnerability varied with different levels of information provided
    }
    \label{tab:GPT-4's analysis of the vulnerability}
    \centering
    \begin{tabular}{c|c|c|c}
    \toprule
    Info  &  Analysis & Repair Strategy & Patch  \\
    \hline
    Type of vulnerability & \cross & \cross &\cross \\
    Crash location & \cross & \cross &\cross \\
    Error message & \tick & \cross &\cross \\
    Crash-Free Constraint  & \tick & \cross &\cross \\
    Error message and type of vulnerability & \tick & \cross &\cross \\
    Error message, type of vulnerability and POC & \tick & \tick &\cross \\
    All of the above & \tick & \tick &\cross \\
    \bottomrule
    \end{tabular}
    \vspace{-10pt}
\end{table}

\subsection{Our approach}

The limited effectiveness of zero-shot repair methods may be attributed, we believe, to a crucial factor: these approaches fail to fully leverage the understanding, reasoning, and dialogic capabilities of LLMs.
Instead, LLMs are utilized in a manner akin to traditional generative models without specific training tailored to the nuances of vulnerability repair, making them struggle with diverse vulnerabilities.
The efficacy of Chain of Thought (CoT) techniques~\citep{wei2022chain, hou2023large, feng2024towards} has been demonstrated, showing that guiding the LLM through structured dialogue improves problem-solving.
Furthermore, extensive research~\citep{zhang2024autocoderover, xia2023keep, bouzenia2024repairagent} has also shown that breaking down a problem into sub-problems and engaging the LLM in ongoing dialogue allows them to deepen their understanding of the overall issue incrementally, resulting in highly effective outcomes.

\begin{figure*}[!th]
  \centering
  \includegraphics[width=\textwidth]{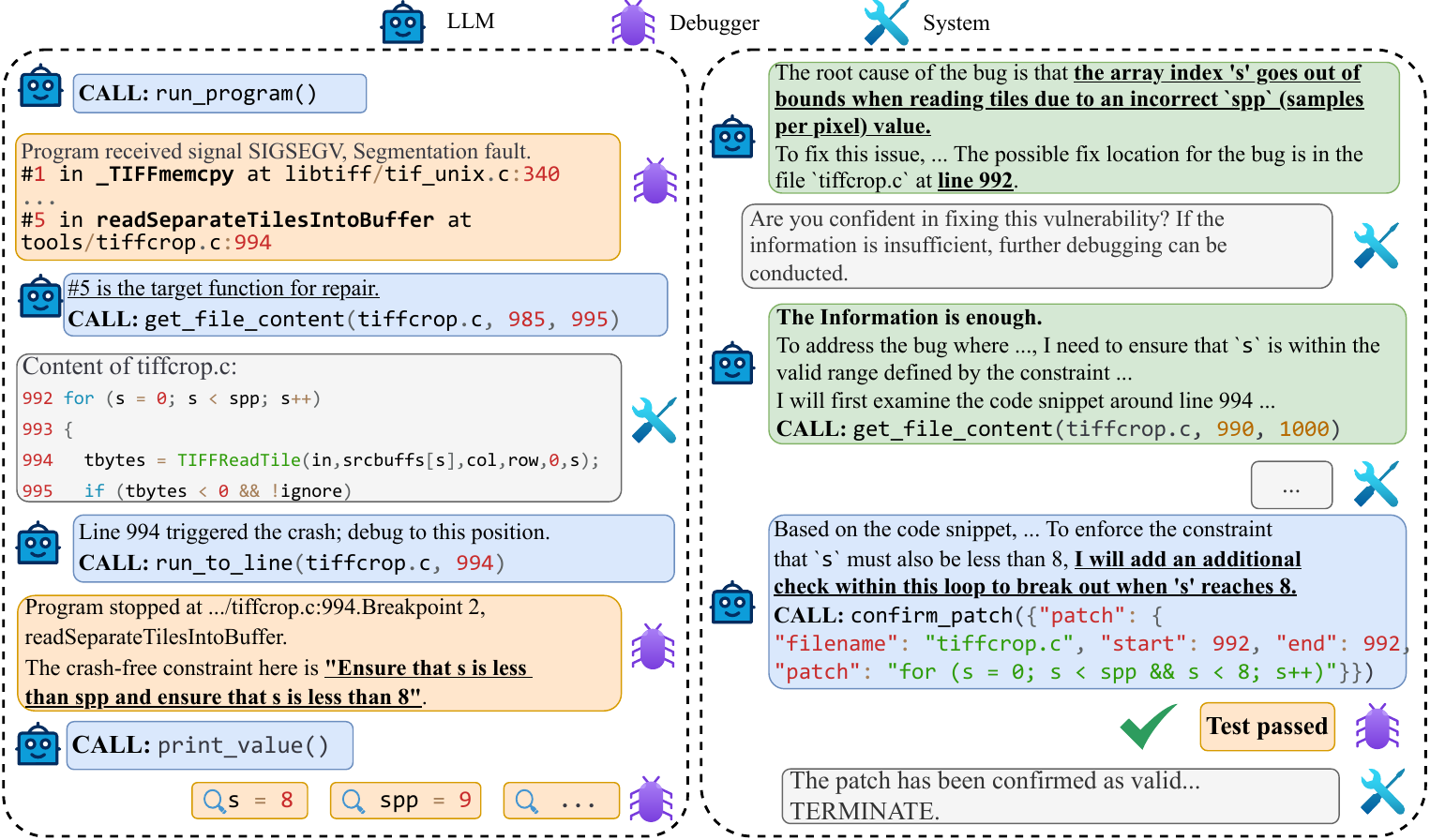}
  \caption{The motivation example}
  \label{fig:motivation}
  \vspace{-6pt}
\end{figure*}


To address the aforementioned issue, we enable the LLM to employ tools like GDB to gather debugging information.
Figure~\ref{fig:motivation} illustrates the debugging and patch generation process of the motivation example using our approach. 
For brevity and clarity, some details and dialogues are omitted.
LLM first calls \texttt{run\_program} to trigger the vulnerability and obtain the crash message and stack trace.
It then identifies the target function and crash line, followed by calling \texttt{get\_file\_content} to retrieve the surrounding code.
After reviewing the code context, the LLM inspects variable values at line 994 using \texttt{run\_to\_line}, setting a breakpoint via the debugger.
This debugging session reveals the actual values at line 994, which are critical: the value of \texttt{spp} is 9, and the value of \texttt{s} is 8. 
Upon learning these real values, we inform the LLM of the crash-free-constraint, i.e., ``Ensure that s is less than spp and ensure that s is less than 8 at line 994'' (the constraint is inferred in ExtractFix~\citep{gao2021beyond}).
Through a comparative analysis of actual and expected states, the LLM gains a deeper understanding of the root cause of the issue. 
The debugging process continues iteratively based on the LLM’s requirements until it obtains sufficient information.
The LLM then revisits the code, come up with a repair strategy, and generates a patch.
Finally, we validate the patch by re-compiling the project to see whether it still triggers the vulnerability or not.

This example shows that our method enables the LLM to debug and repair like a human.
The LLM accesses contextual and debugging information via the APIs provided by the toolset, continually advancing its understanding of the root causes of vulnerabilities through the analysis of the expected and actual states.
This approach fully leverages the LLM's capabilities in understanding and reasoning, effectively overcoming zero-shot limitations.


\begin{figure*}[!th]
  \vspace{-6pt}
  \centering
  \includegraphics[width=\textwidth]{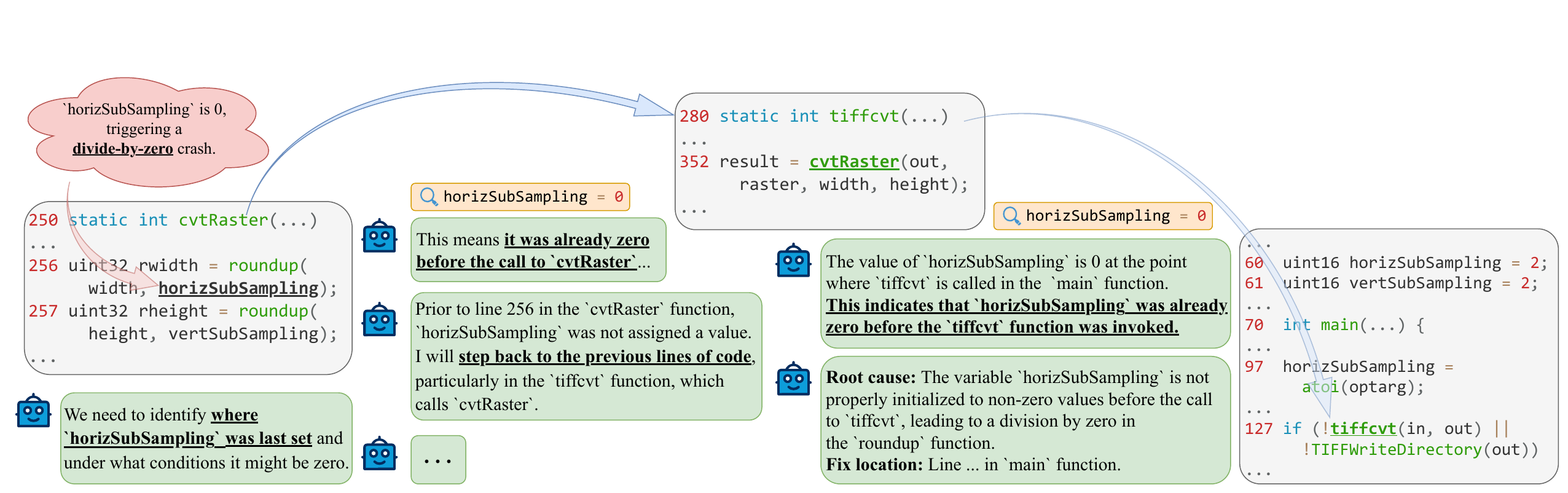}
  \caption{The debugging of CVE-2016-3623}
  \label{fig:motivation2}
  \vspace{-6pt}
\end{figure*}

Figure~\ref{fig:motivation2} presents another example that better illustrates the debugging process. 
The vulnerability, identified as CVE-2016-3623~\citep{cve-2016-3623}, is depicted in line 256.
Here, a divide-by-zero crash is triggered due to \texttt{horizSubSampling} being set to 0. 
This value originates from an assignment in \texttt{main} at line 97, which then calls \texttt{tiffcvt} (line 127) and \texttt{cvtRaster} (line 352), where the crash occurs.
After obtaining the preliminary crash information, the LLM initially confirms a debugging strategy based on the constraint that $\mathtt{horizSubSampling} \neq 0$, with LLM stating ``we need to identify where \texttt{horizSubSampling} was last set and under what conditions it might be zero.''
Accordingly, LLM first invokes \texttt{get\_file\_content} to retrieve the relevant code lines within \texttt{cvtRaster}.
Upon confirming that there are no assignments to \texttt{horizSubSampling} in \texttt{cvtRaster}, the LLM then moves to the prior stack frame to check the last invocation of \texttt{cvtraster}, which is within \texttt{tiffcvt}, revealing \texttt{horizSubSampling} is already zero. 

Consequently, further examination of the relevant code lines within \texttt{tiffcvt} confirms that this function had no assignments to \texttt{horizSubSampling} either. Following the same procedure, LLM then goes to line 127 in \texttt{main}. 
The debugging information still shows that \texttt{horizSubSampling} is zero.
After revisiting the context, LLM decided to implement a fix at line 126. 

In fact, fixing a divide-by-zero vulnerability is not challenging, which merely requires preventing that $\mathtt{horizSubSampling}$ equals zero before the bug is triggered.
Therefore, alternative repair methods often involve inserting a conditional statement near line 255.
However, the theoretically optimal location for fixing a divide-by-zero error is immediately after the erroneously zeroed variable is assigned the value of zero.
It is ideal to interrupt or reassign the erroneous operation as soon as possible to prevent unexpected program behavior.
Our approach utilizes the constraint $\mathtt{horizSubSampling} \neq 0$ to guide LLM through the debugging process, enabling LLM to acquire dynamic information at various points, confirm the actual location where $\mathtt{horizSubSampling}$ is set to zero, and subsequently carry out the repair closer to the optimal location.
This is a process that existing methods struggle to achieve.

\section{Methodology}

In this section, we present the design of \toolname, and describe the detailed approach for utilizing both static and dynamic contexts to guide the LLM in repairing vulnerabilities.

\begin{figure*}[h]
  \centering
  \includegraphics[width=\textwidth]{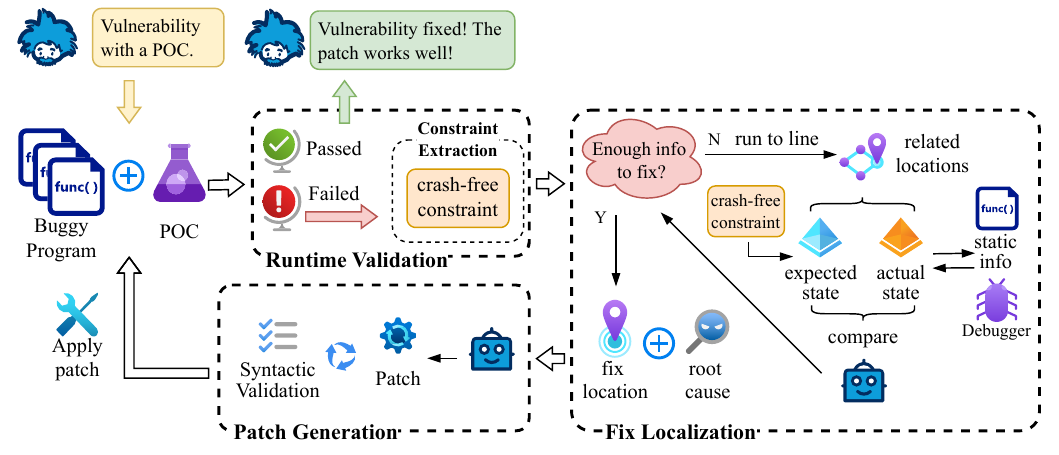}
  \vspace{-12pt}
  \caption{Overall framework of \toolname}
  \label{fig:framework}
  \vspace{-6pt}
\end{figure*}

Figure~\ref{fig:framework} shows the overall workflow of the proposed technique. 
First, \toolname runs the program with the POC that triggers the vulnerability to cause a crash.
With the crash information and extracted crash-free constraint, \toolname constructs the initial prompt, and commences the debugging process.
Throughout this process, \toolname provides the LLM with a suite of APIs to access the static context of the program and obtain dynamic information.
With the crash-free constraint in ``mind'', the LLM sets breakpoints at various locations within the program and conducts debugging to acquire the necessary dynamic context.
Once the problem is well analyzed, the LLM outputs the root cause of the vulnerability and possible fix locations.
Based on this information, \toolname again guides the LLM to generate patches, and validate them by trying to reproduce the vulnerability.
Patches that do not trigger the crash are finally output as the repair results.

\subsection{Perceiving the expected state based on the crash-free constraint}
\label{sec:perceive}

We utilize the crash-free constraint (CFC) as the expected state for the crash location and a basis for inferring the expected states of other locations. 
In this section, we will explain how we extract the CFC and perceive the expected states in potential fix locations.


\subsubsection{Crash-free constraint extraction}

Gao et al.~\citep{gao2021beyond} proposed a vulnerability repair methodology named \ExtractFix, based on extracting CFCs to avoid patch overfitting.
This approach extracts a constraint representing the vulnerability with the aid of sanitizers, which then serves as the basis for synthesizing patches.
Eventually, this method represents CFCs through predefined templates.
However, LLMs may struggle to accurately comprehend expressions that have not appeared in their training set.

For example, Listing~\ref{GNUBug25003} demonstrates a patch for the vulnerability identified as GNU Bug 25003~\citep{gnu_bug_25003}.
Prior to the correction, the conditional check \texttt{if (initial\_read != SIZE\_MAX || start < initial\_read)} would not proceed to evaluate \texttt{start < initial\_read} if \texttt{initial\_read != SIZE\_MAX} was true.
Consequently, this led to the invocation of \texttt{memmove(buf, buf + start, initial\_read - start)} with the third parameter potentially being less than zero.
Given that this parameter is of type unsigned integer, a negative value passed to \texttt{memmove} would be interpreted as a substantially large positive one.
This situation would result in \texttt{memmove} attempting to shift a much larger memory block than intended or available, leading to buffer overflow.

\begin{lstlisting}[xleftmargin=23pt, language=C, caption={GNU Bug 25003}, label={GNUBug25003}]
- if (initial_read != SIZE_MAX || start < initial_read) {
+ if (start < initial_read) {
    memmove (buf, buf + start, initial_read - start);
    initial_read -= start;
}
\end{lstlisting}

For the vulnerability in question, Listing~\ref{constraint} shows the constraint extracted by \ExtractFix.
The constraint simplifies to \texttt{start < initial\_read}, accurately representing the required CFC.

\begin{lstlisting}[xleftmargin=23pt, language=C, caption={The constraint extracted by \ExtractFix}, label={constraint}, escapechar=|]
(And (Or (Not (Eq false
                   (Eq 18446744073709551615 initial_read)))
          (Ule 0 (Sub  initial_read start)))
      (Or (Not (And (Eq 18446744073709551615 initial_read)
                    (Ult  start 18446744073709551615)))
          (Ule 0 (Sub  initial_read start))))
\end{lstlisting}

However, GPT-4 misinterpreted the expression 
\texttt{Not (Eq false (Eq 18446744073709\\551615 initial\_read))} as \texttt{initial\_read $\neq$ SIZE\_MAX}.
The correct simplification should have been \texttt{initial\_read == SIZE\_MAX}. 
This incorrect interpretation led to a flawed analysis, causing GPT-4 to conclude erroneously that the pre-repaired code snippet meets the constraints and therefore has no vulnerabilities.
Such an error can lead to a completely incorrect repair process.

To address the aforementioned issues, we simplify the constraint expressions extracted by \ExtractFix and design a straightforward template to convert them into natural language descriptions.
Table~\ref{tab:constraints} shows how various CFC templates are translated into natural language.
Besides the shown expressions, templates for logical and arithmetic operators are also included.
Using this framework, the constraints are transformed into natural language such as ``\textit{Variable start should be less than variable initial\_read}'', helping large models better understand CFCs for more accurate program state assessments.

\begin{table}[th]
    \footnotesize
    \caption{CFC templates and their corresponding natural language templates}
    \vspace{-6pt}
    \label{tab:constraints}
    \centering
    \begin{tabular}{c|c|c|m{2.5cm}|m{5.5cm}}
    \toprule
    Class  &  ID & Expression & \makecell{\centering CFC Template} & \makecell{\centering Natural Language Template} \\
    \hline
    Developer & $T_1$  &  \texttt{assert(C)} & \centering \texttt{C} & Ensure that <ConditionDesc>. \\ \hline
    
    \multirow{5}{*}[-2em]{Sanitizer} & $T_2$ & \texttt{*p} & \texttt{p + sizeof(*p) \(\leq\) base(p) + size(p) \(\land\) p \(\geq\) base(p)} & Pointer <Pointer> should be within its allocated bounds \\
    \cline{2-5}
    & $T_3$ & \texttt{a $op$ b} & \texttt{MIN $\leq$ a $op$ b $\leq$ MAX} & The result of <Variable> <Operator> <Variable> should be within the range from <Number> to <Number> \\
    \cline{2-5}
    & $T_4$ & \shortstack{\texttt{memcpy(p, q, s)}} & \texttt{p + s $\leq$ q $\lor$ q + s $\leq$ p} & The memory regions defined by <Variable> and <Variable> should not overlap \\
    \cline{2-5}
    & $T_5$ & \texttt{*p} & \centering \texttt{p $\neq$ 0} & Pointer <Pointer> should points to a valid address \\  \cline{2-5}
    & $T_6$ & \texttt{a / b} & \centering \texttt{b $\neq$ 0} &  Variable <Variable> should not be equal to zero \\   
    \bottomrule
    \end{tabular}
    \vspace{-6pt}
\end{table}

\subsubsection{Expected states in potential fix locations.}

Utilizing CFCs, we capture the expected state of the program at the point where a crash is triggered. 
However, human developers will anticipate a specific state at each breakpoint.
\ExtractFix employs forward symbolic execution to propagate constraints, yet obtaining expected states via constraint propagation is not always effective.

For example, during one of our trials with the example in Figure~\ref{fig:motivation}, LLM attempted to obtain the value of \texttt{row}.
This operation may appear unrelated to the CFC and the crash itself, yet it aids in understanding the overall context and the nesting levels of loops.
However, propagating CFCs often fails to provide such anticipated states.
Moreover, the overhead associated with performing symbolic execution to propagate constraints each time LLM identifies a debugging target is also unacceptable.

Therefore, the lightweight guidance for LLM, based on CoT techniques, may exhibit superior performance in addressing this issue.
Specifically, after each selection of a debugging target line by LLM, \toolname uses the prompt ``\textit{Think of the constraint and expected state of the program here based on the crash-free constraint. Compare it with the real state of the program to deepen the understanding of the bug.}''
This guides LLM in contemplating the significance of the current debugging effort.
This process is designed to enhance its logical reasoning and reduce the occurrence of irrelevant and non-productive debugging activities.

\subsection{Obtaining actual state through program debugging}

After discussing how we ascertain the program's expected state through crash constraints, this section will detail how we utilize LLM to obtain dynamic program information through program debugging.
To achieve an automated debugging process, we provide LLM with the following two categories of APIs.

\begin{itemize}[leftmargin=*]
    \item \textbf{Static information retrieval.}
    Due to LLM's inherited input constraints, feeding an entire project-level source code directly into an LLM for analysis is impractical. 
    However, source code access is crucial for debugging.
    To address this challenge, similar to existing methods~\citep{zhang2024autocoderover}, we enable the LLM to autonomously access source code by providing the following APIs.
    \begin{itemize}[leftmargin=*]
        \item \textit{definition} Get the definition of a symbol in the code.
        \item \textit{summary} Retrieve the signature and related comments of a symbol (e.g., function or variable).
        \item \textit{function\_body:} Retrieve the complete definition of a function.
        \item \textit{get\_file\_content}  Get the content of a file in the given range.
    \end{itemize}
    \item \textbf{Dynamic information retrieval.} 
    To facilitate the process of debugging programs using LLM, we have provided a suite of program debugging APIs listed below.
    \begin{itemize}[leftmargin=*]
        \item \textit{run\_program} Run the program in debugger and return the error message and backtrace.
        \item \textit{run\_to\_line}
        Debug the program until the specified line to retrieve the actual state.
        \item \textit{print\_value}
        Get the actual value of a variable or expression in the current context.
    \end{itemize} 
\end{itemize} 

\subsubsection{Debugging process based on toolset}

\begin{figure*}[!th]
  \centering
  \includegraphics[width=0.9\textwidth]{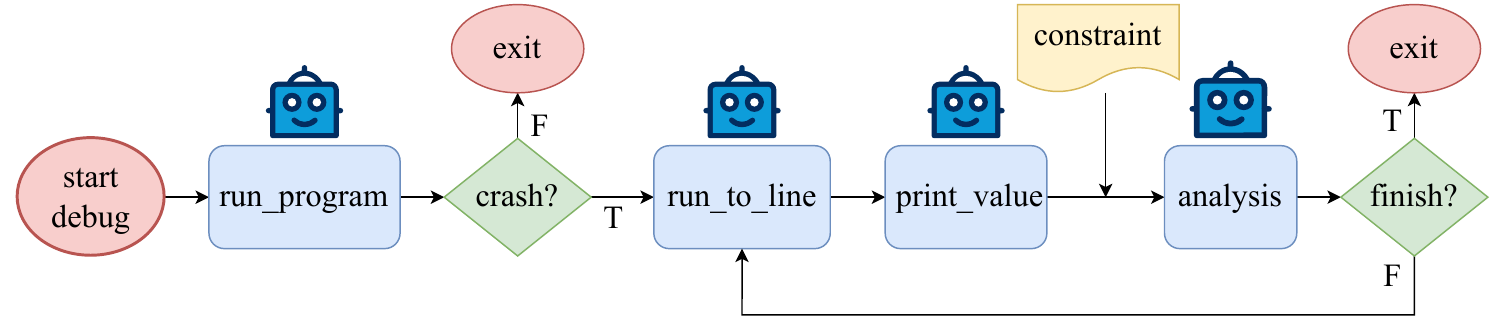}
  \vspace{-6pt}
  \caption{Debugging process of \toolname}
  \label{fig:debug}
  \vspace{-6pt}
\end{figure*}

Figure~\ref{fig:debug} illustrates the process of using LLM for debugging to ascertain the actual state of the program.
Initially, \toolname invokes \texttt{run\_program} to conduct the first run of the program using debugging tools.
This aims to trigger the crash, thereby obtaining the initial error information that guides the subsequent debugging.
Based on the stack trace provided by the error, LLM identifies suspicious functions that could cause the crash, and conducts debugging sessions for each one.
In a debug session, the LLM invokes \texttt{run\_to\_line}, which sets a breakpoint within the target frame to pause the program at a specific line in the suspicious function, making the actual state of the program here accessible.
As described in Section~\ref{sec:perceive}, with the preceived expected state, LLM retrieves related static information, and invokes \texttt{print\_value} to inspect the specific variable or expression information.

\subsubsection{Static information retrieval based on toolset}

In addition to using debugging techniques to understand vulnerabilities, we have also explored how static program information can assist and enhance the debugging process.
Source code and explicit line numbers are crucial for the LLM to select a valid breakpoint location, therefore, we provide \texttt{get\_file\_content} to retrieve code segments with line numbers added at the start of each line.
Furthermore, it is hard to tell the type of a symbol by its name only, so we provide \texttt{definition} to retrieve a symbol's definition statement.
It is sufficient in most cases, but there could be type aliases, or functions that has obscure parameter names, making this definition statement useless.
However, the good news is that developers often kindly leave some notes or documentation in comments that explain everything.
Considering this, we have \texttt{summary} to provide an analyzed and formatted summary for a symbol by resolving all involved type aliases and fetching its surrounding comments.
Additionally, we provide the \texttt{function\_body} API to fetch the entire source code of a function used in the context.

\subsection{Patch generation and validation}

In this section, we will explain how \toolname utilizes information obtained from the debugging process to generate patches.
Additionally, we will discuss our method for performing a preliminary validation of the patches.

\subsubsection{Summary of debugging information}

Once the LLM deems that it has gathered sufficient information during the degugging process, we guide the LLM to summarize it to get root cause and possible fix location.
The summary of our debugging information, denoted as $S$, can be defined as follows.

\begin{equation}
    S = \bigoplus_{i=1}^{n} c(\psi_i, \Gamma_i) \vdash r, l
\end{equation}

In this formula, $n$ represents the total number of debugging iterations. 
$\psi_i$ denotes the expected state predicted during the $i$-th debugging iteration, while $\Gamma_i$ denotes the actual state predicted at the same iteration.
The operator $\bigoplus$ symbolizes the result of LLM's comparison between them.
Furthermore, $r$ and $l$ represent the root cause and fix location, respectively.
This formula indicates that the summary of debugging information comprises two components: a detailed root cause analysis of the vulnerability contextualized within the reproducing environment and a precise determination of the repair location. 
Drawing upon established expertise in vulnerability detection and repair~\citep{dissanayaka2020vulnerability, zhou2012should}, we posit that to repair a vulnerability, it is essential to understand the root cause and achieve precise repair localization.
The last LLM dialogue box in Figure~\ref{fig:motivation2} presents an example of a genuine root cause analysis.
Guided by our directions, the LLM provides a highly specific and comprehensive analysis of the root cause, which effectively aids in guiding the patch generation process.

\subsubsection{Patch generation and validation}

\begin{figure*}[!th]
  \centering
  \includegraphics[width=0.9\textwidth]{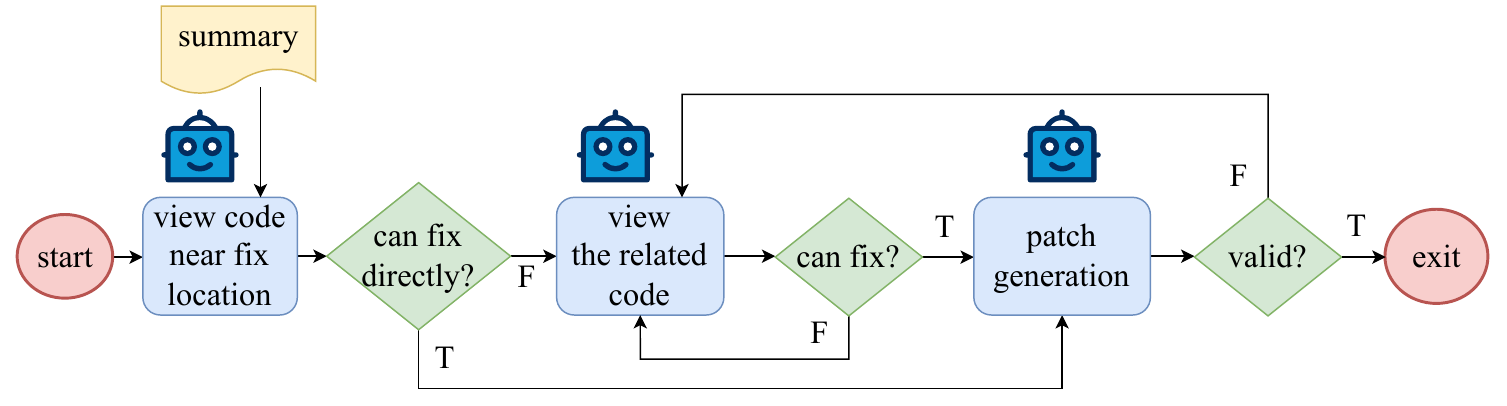}
  \caption{Patch generation process of \toolname}
  \label{fig:pg}
  \vspace{-12pt}
\end{figure*}

Patch generation is handled by a separate agent with a fresh conversation to avoid exceeding token limits from the extensive debugging dialogue.
Utilizing a new agent for patch generation allows the LLM to focus more specifically on this task while simultaneously reducing the interference from redundant information. 

Figure~\ref{fig:pg} illustrates the patch generation process, taking the summarized root causes and repair locations from the previous step as input.
Initially, the LLM reviews the code at the repair location and assesses whether a direct repair is feasible.
If direct repair is not possible, potentially due to erroneous localization, the LLM then performs a secondary localization based on the root cause.
During this process, the LLM continuously examines code segments involving relevant variables and functions, sequentially evaluating their reparability.
If the LLM determines that a repair is feasible, it generates a patch for \toolname to validate.
\toolname then checks for simple syntax errors, such as mismatched parentheses, and replaces the patch in the source code before re-executing the vulnerability exploit. 
\toolname then checks basic syntactic and semantic errors by compiling the patched project and re-running it to see if the vulnerability persists.
If the exploit no longer triggers the crash, the process concludes.
Otherwise, \toolname will conduct another iteration for other possible locations to fix.


\section{Implementation and Evaluation}

In this section, we aim to evaluate \toolname and answer the following research questions:

\begin{itemize}[leftmargin=*]
    \item \textbf{RQ1:} Compared to state-of-the-art approaches, how effective is \toolname in repairing vulnerabilities?
    \item \textbf{RQ2:} How effective is the debugging process?
    \item \textbf{RQ3:} What impact do the crash-free constraints have on repair precision?
\end{itemize}

\subsection{Implemetation}

We employ the open-source framework AutoGen~\citep{AutoGen2023} to construct our agent.
It allows LLM to invoke external tools and provide seamless support for different LLM models.
We mainly utilize OpenAI's GPT-4o as the foundational reasoning model for \toolname that balances cost and effectiveness.
For parameters, we set a low temperature of 0.2 to produce relatively deterministic results, and other parameters remain as per default.

For a generated patch, \toolname will validate it by trying to reproduce the vulnerability with the patched project.
Failures trigger an LLM prompt with an error description for patch regeneration as hint for retry.
The number of permissible failures before exiting the patch generation cycle is set to three.
If all attempts fail, LLM retries debugging to refine fix localization and root cause analysis.

Based on our choice for the dataset, Clang is used as it supports more sanitizer and fuzzing options.
To be more compatible with the compiler, we use LLDB 10.0.0 as the debugger and a custom build of LLDB's machine interface driver.
For static information, we utilize clangd as the Language Server.
However, our methodology is theoretically not limited by the programming language.


\subsection{Experimental Dataset}

To evaluate the effectiveness of \toolname, we use Juliet test suites benchmark~\citep{black2018juliet, NIST_SARD_112}, and real-life projects from \ExtractFix~\citep{ExtractFix2023} and ARVO~\citep{mei2024arvoatlasreproduciblevulnerabilities}.
The Juliet test suites benchmark is a collection of test cases containing 64099 examples organized under 188 different CWEs, with each case consists of one buggy file and a brief description.
\ExtractFix comes with 32 cases with constraints extracted already.
The ARVO dataset contains 5,001 reproducible vulnerabilities across 273 projects detected by OSS-Fuzz.
All these cases can be reproduced by applying specific sanitizer options.
To apply our method, we filter these datasets based on the following criteria.

\begin{itemize}[leftmargin=*]
    \item \textbf{Programming Language}
    Although C and C++ are often addressed together in program repair tasks, we currently focus on C projects to avoid the complex debugging context of C++.
    \item \textbf{Constraint Availability}
    Constraints can be successfully extracted for the target project.
    \item \textbf{Environment}
    The target vulnerability can be reproduced in Ubuntu 20.04, which is required by our custom build of lldb-mi.
    \item \textbf{Compilation Overhead}
    The complete workflow of \toolname involves multiple compilations of the target project, especially for patch validation. Therefore, we require the target project to be compiled within two minutes to ensure a reasonable time.
\end{itemize} 

Although ARVO has thousands of cases, more than 60\% of them were discovered before 2022 and are reproduced within Ubuntu 16, thereby cannot be used by us directly.
And for the rest, many are C++ projects, or we are unable to extract constraints from them.
Eventually, we picked 25 representative cases from Juliet test suites, 14 projects from \ExtractFix, and 36 projects from ARVO.

\begin{table*}[!t]
    \centering
    \captionsetup{justification=centering}
    \caption{Comparison of repair results \\ SE: Semantically Equivalent, P: Plausible}
    \begin{tabular}{c|c|cccc}
        \toprule
        \multicolumn{1}{c|}{\multirow{2}{*}{Benchmark}} & \multirow{2}{*}{\#Vul}
        &\multicolumn{4}{c}{Repair Results (SE/P)} \\ 
        \cline{3-6}
        & & \VulRepair & LLM & \autocoderover & \toolname \\
        \hline \hline
        \multicolumn{6}{c}{Results on real-life projects} \\
        \hline
        gpac         & 10  & 0/0 & 0/1 & 0/4 & 4/3 \\
        libxml2      &  9  & 0/0 & 0/4 & 0/2 & 3/2 \\
        libjpeg      &  5  & 0/1 & 1/0 & 0/1 & 1/2 \\
        libtiff      &  4  & 0/1 & 1/0 & 2/0 & 2/2 \\
        libplist     &  4  & 0/0 & 0/0 & 0/0 & 1/1 \\
        file         &  4  & 0/0 & 1/0 & 0/2 & 3/1 \\
        ndpi         &  4  & 0/0 & 0/0 & 2/0 & 2/0 \\
        jasper       &  2  & 0/0 & 0/0 & 0/0 & 1/0 \\
        elfutils     &  2  & 0/0 & 0/0 & 0/1 & 1/0 \\
        htslib       &  2  & 0/0 & 0/0 & 0/0 & 0/0 \\
        libcoap      &  1  & 0/0 & 0/0 & 0/0 & 0/0 \\
        cups         &  1  & 0/0 & 0/0 & 0/0 & 0/0 \\
        cyclonedds   &  1  & 0/0 & 0/0 & 0/0 & 0/0 \\
        lcms         &  1  & 0/0 & 0/0 & 0/0 & 1/0 \\
        \hline
        Total        & \multirow{2}{*}{50} & 0/2     & 3/5     & 4/10    & 19/11    \\
        \cline{1-1} \cline{3-6}
        Precision    &                     & 4.00\%  & 16.00\% & 28.00\% & 60.00\%  \\ 
        \hline \hline
        \multicolumn{6}{c}{Results on Juliet test suites} \\
        \hline
        Juliet       & \multirow{2}{*}{25} & 4/1     & 18/2    & -       & 22/2     \\
        \cline{1-1} \cline{3-6}
        Pecision     &                     & 20.00\% & 80.00\% & -       & 96.00\%  \\
        \bottomrule
    \end{tabular}
    \label{tab:overall results}
    \vspace{-6pt}
\end{table*}

\subsection{RQ1: Effectiveness of \toolname}

To answer this question, we evaluate the success rate of \toolname and compare it with existing tools.

We choose \VulRepair~\citep{fu2022vulrepair}, \autocoderover~\citep{zhang2024autocoderover}, and the LLM as our comparison tools. \VulRepair is a T5-based automated vulnerability repair approach. 
\autocoderover is an agent-based repair tool with abilities that exploits program structure by code searching.
Due to the lack of code repositories and issue information in the Juliet test suites, \autocoderover is excluded from that benchmark.
To address potential data leakage concerns, we also compared LLM-generated patches using conversation-only interactions.
We argue that if a vulnerability is not directly fixed by LLM but is successfully addressed by \toolname, the result is unrelated to data leakage.
In this study, we provided LLM with the function body that contained vulnerabilities and crash information, allowing it to generate five patches.
We then used the patch closest to being correct as the basis for our statistical results.
The LLM model configuration used in \autocoderover and LLM is identical to that of \toolname.

To evaluate the effectiveness of the above tools, we have assessed their repair accuracy.
We categorized the patches into three levels: \textit{fail}, \textit{plausible}, and \textit{semantically equivalent}.
In this context, \textit{fail} refers to the inability to generate a patch that passes the POC test.
\textit{plausible} denotes patches that pass but deviate from the developer's logic.
\textit{Semantically equivalent} indicates patches that are semantically identical to the original.
Ultimately, the tool's vulnerability repair precision is determined by the ratio of the sum of vulnerabilities classified as \textit{semantically equivalent} and \textit{plausible} to the total number of vulnerabilities.
To mitigate randomness, we use pass@3~\citep{chen2021evaluating} as the evaluation metric. All LLM-related experiments, including the comparative analysis with other tools, adhere to this principle.


In terms of CFCs, for the projects from \ExtractFix dataset, we directly applied the CFCs provided by \ExtractFix.
For those in the Juliet test suite and ARVO dataset, we extract CFCs following the methodology of \ExtractFix.



\textit{\textbf{Results.}} 
Table~\ref{tab:overall results} summarizes the results of \VulRepair, \autocoderover, LLM, and \toolname{}.
For large-scale real-world projects, \toolname achieved 60\% accuracy with 19 \textit{semantically equivalent} and 11 \textit{plausible} patches.
In contrast, \VulRepair generated 2 \textit{plausible} patches (4\% accuracy), while \autocoderover produced 4 \textit{semantically equivalent} and 10 \textit{plausible} patches (28\% accuracy).
The LLM achieved 16\% accuracy with 3 \textit{semantically equivalent} and 5 \textit{plausible} patches.
On the Juliet test suites benchmark, \toolname generated 22 \textit{semantically equivalent} patches and 2 \textit{plausible} patches, with 1 patch failing the POC.
In comparison, \VulRepair completed 4 \textit{semantically equivalent} patches and 1 \textit{plausible} patch.
The LLM produced 18 \textit{semantically equivalent} patches and 2 \textit{plausible} patches. 

The experimental results show that our dynamic state-aware agent can effectively repair vulnerabilities. This is because the dynamic information available during program execution enhances the LLM's understanding of the process that triggers the error. Furthermore, by comparing this with the expected states suggested by the CFC, the LLM's comprehension of the fundamental causes of the vulnerabilities is further strengthened.

Although \toolname performs well in terms of averaged precision, it fails to repair certain vulnerabilities.
For instance, the vulnerability CVE-2016-10094~\citep{cve-2016-10094} in Libtiff~\citep{libtiff} is a heap overflow, occurring at the line \texttt{\_TIFFmemcpy(buffer, jpt, count - 2);} when \texttt{count} equals 4, leading to a program crash.
After debugging, \toolname identified the root cause: \textit{``TIFFGetField'' did not set ``count'' and ``jpt'' correctly, leading to invalid memory access. The condition ``if (count >= 4)'' is not sufficient to ensure ``jpt'' points to valid memory}.
The root cause analysis in terms of condition evaluation is accurate, and the developer's patch modified line 2898 from \texttt{if (count >= 4)} to \texttt{if (count > 4)}.
However, the focus on the variables and the understanding of the error were incorrect. 
At the crash, \texttt{jdt} did not point to invalid memory access, misleading subsequent repairs.
The issue arose during debugging when \toolname incorrectly set a breakpoint on an if statement, causing the program to stop at its first encounter,
However, the crash was not triggered immediately after this first encounter, but a subsequent one.
As a result, the captured state did not match LLM’s expectations, leading to the mistaken assumption that the variable \texttt{jdt} was uninitialized, and thus, an incorrect root cause was identified.

\subsection{RQ2: Effectiveness of the debugging process}

\subsubsection{Effectiveness of the root cause analysis and fix localization.}

The root cause and fix location summarize debugging results and guide patch generation.
Their accuracy reflects how well debugging improves LLM's code understanding and directly impacts patch correctness.
With the intermediate results of the real-life projects from RQ1, we analyzed how the accuracy of root cause and fix location contributes to the precision of patch generation.
A root cause \(r\) is considered correct if its suggestion is directly applied to generate a patch and identifies the specific cause of the vulnerability.
Similarly, a fix location \(l\) is correct if we can generate patches at \(l\) that are semantically equivalent to the developer's patches.


\begin{wraptable}{r}{0.6\textwidth}
    \centering
    \caption{Impact of root cause and fix location} 
    \vspace{-6pt}
    \begin{tabular}{c|ccc}
        \toprule
         Debug Result                              & \#Vul & Fixed & Precision \\
        \hline
         Both \(r\) and \(l\) are correct          & 33    & 25    & 75.76\%   \\
         One of \(r\) and \(l\) is correct         & 6     & 3     & 50.00\%   \\
         None of \(r\) and \(l\) is correct        & 7     & 2     & 28.57\%   \\
         Failed to provide \(r\) or \(l\)          & 4     & 0     & 0.00\%    \\
        \bottomrule
    \end{tabular}
    \label{tab:r and l}
    \vspace{-6pt}
\end{wraptable}

\textit{\textbf{Results.}} As is shown in Table~\ref{tab:r and l}, \toolname can successfully provide correct root causes and fix locations for 33 out of 50 projects. 
In this case, it raises the precision of the generated patch to 75.76\%.
These figures indicate that \toolname can achieve highly accurate repairs with minimal overhead.
Respectively, the precision drops if the \toolname fails to reason the correct root cause or fix location.
This indicates that the overall precision of \toolname relies on more effective debugging results, which positively contribute to higher quality of the generated patches. 
The detailed results for each case are presented in the appendix.


\subsubsection{Study on the rounds of debugging.}

\begin{figure}[h]
\vspace{-6pt}
  \centering
  \includegraphics[width=0.8\textwidth]{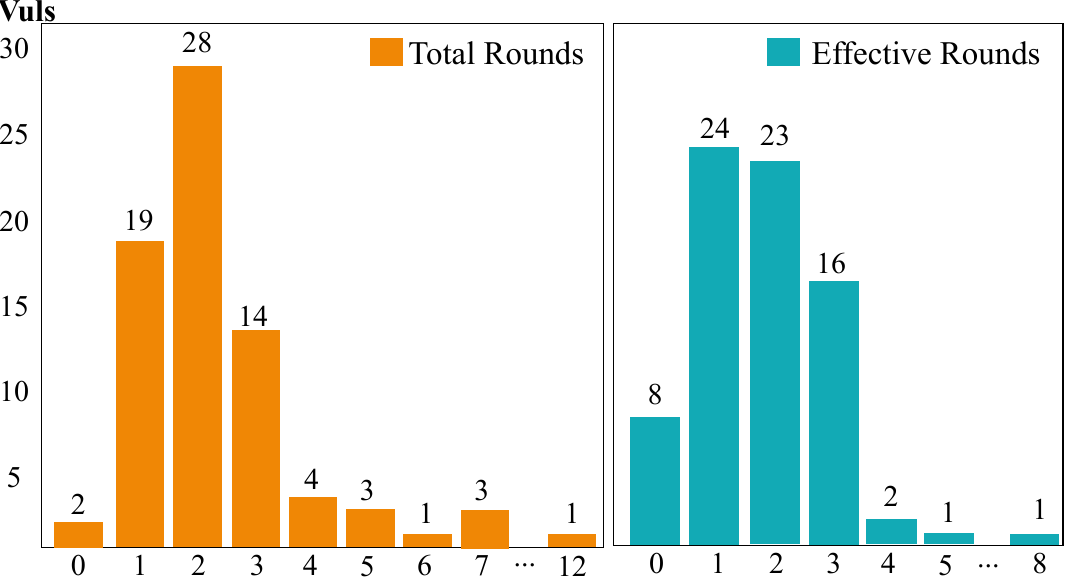}
  \vspace{-6pt}
  \caption{Degugging rounds}
  \label{fig:debug rounds}
  \vspace{-6pt}
\end{figure}

We further collected statistics on the debugging rounds required to repair each vulnerability and the number of effective rounds.
Debugging rounds were counted based on LLM calls to the \texttt{run\_to\_line} API, excluding the mandatory crash-site debugging session.
We consider a debugging session to be effective if it meets at least one of the following two conditions:
(1) Examines key crash-related variables or analyzes variables with explanations.
(2) Reviews relevant code snippets used in subsequent debugging or localization fixes.
The experimental setup follows \toolname in RQ1. The analysis focuses on the debugging process during the initial vulnerability repairs performed by \toolname.

\textit{\textbf{Results.}} Figure~\ref{fig:debug rounds} shows the debugging round statistics. 
The x-axis represents the debugging round \(n\), while the y-axis indicates the number of patches given at \(n\).
The left figure shows the total number of rounds, while the right figure presents the count of effective rounds.
The experimental data reveal that for the majority of vulnerabilities, \toolname requires 2 debugging sessions to analyze the root cause. 
Except for 1 outlier, the number of debugging rounds does not exceed 7. 
Furthermore, the 2 instances where the number of debugging sessions was 0 are due to the root cause being straightforward, with debugging at the crash site providing sufficient information. 
The 12-session case involved analyzing a complex structure, requiring multiple iterations to capture attribute values.
The statistics for effective debugging sessions reveal that 8 repairs had none, 24 patches had 1 effective session, 23 patches had 2 sessions, and 16 patches required 3 sessions.
The experimental results demonstrate that our method indeed provides additional information through dynamic context and is capable of guiding or assisting LLM in the repair of vulnerabilities.


\subsection{RQ3: Impact of crash-free constraint}

\begin{wraptable}{r}{0.4\textwidth}
\centering
\vspace{-12pt}
\caption{Impact of crash-free constraint}
\begin{tabular}{cc}
    \toprule
     Tool & Real-life Projects \\
    \midrule
    \toolname        & 30/50 (60.00\%) \\
    \toolname$^c$    & 22/50 (44.00\%) \\
    \bottomrule
\end{tabular}
\label{tab:cfc}
\end{wraptable}

In the preceding sections, we discussed perceiving the expected state based on the CFC.
To evaluate the effectiveness of this process, we evaluated \toolname without CFCs.
We eliminated the content related to CFC and the CoT prompts that guide the underlying LLM in perceiving the expected state.
All other experimental configurations remain the same as in RQ1.

\begin{wrapfigure}{r}{0.4\textwidth}
    \centering
    \vspace{-12pt}
    \includegraphics[width=0.32\textwidth]{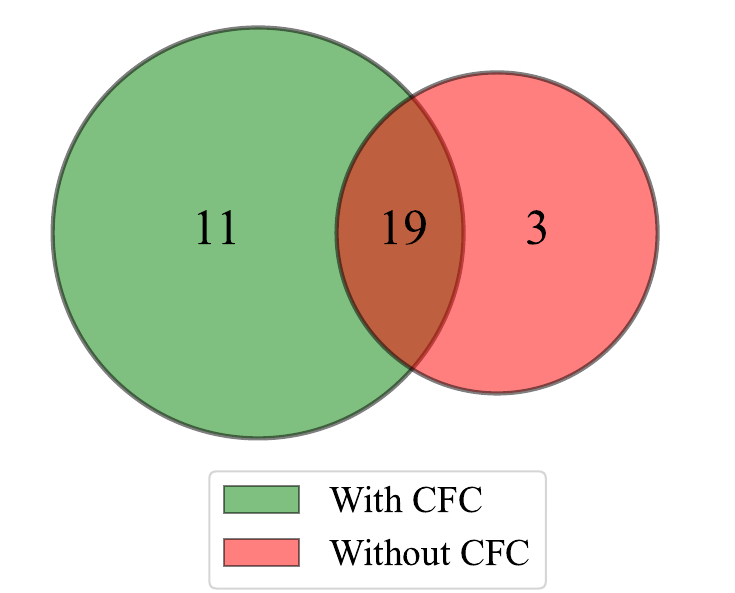}
    \caption{The Venn diagram of fixed vulnerabilities with/without crash-free constraints}
    \label{fig:cfc-venn}
    \vspace{-12pt}
\end{wrapfigure}

\textit{\textbf{Results.}}
As shown in Table~\ref{tab:overall results}, for real-life projects, \toolname can fix a total of 30 vulnerabilities out of 50.
However, according to our statistics, if constraints were not given, then \toolname only managed to fix 22 of them.
We can see that, with the help of CFCs, \toolname successfully generated 8 more patches, indicating that CFCs can indeed enhance \toolname's capability to repair vulnerabilities.
This improvement is attributed to CFC directly reflecting the program's expected state at the crash site or providing hints, giving \toolname more effective information.

Interestingly, as seen in Figure~\ref{fig:cfc-venn}, we noticed that 3 vulnerabilities which \toolname failed to fix with CFCs were successfully patched when CFCs were not provided.
This is because the CFCs can imply a message that the related variables are crucial to fix the vulnerability, so when the LLM is unable to inspect their values, it may not be confident enough to give a root cause or other analysis.
This occurs when the specified variable is optimized out by the compiler, becoming inaccessible for the debugger even with the debug option enabled.
So in this case, the LLM may simply give up, or try other variables and locations, thus leading to incorrect results.

\section{Discussion}

In this section, we discuss the limitation of \toolname and the threats that may affect the validity of our evaluation.

\textit{Limitations.} 
Firstly, \toolname requires a POC to perform vulnerability remediation. 
However, obtaining a POC can be challenging, particularly for vulnerabilities identified by detection tools.
This reliance on POCs limits \toolname's capability to address zero-day vulnerabilities effectively.
Second, while the CFC provides crucial expected state information for debugging processes, its fundamental reliance on program crashes introduces significant limitations.
This dependency renders \toolname ineffective in generating patches for a broader spectrum of software errors, particularly restricting its repair capabilities to crash-inducing vulnerabilities rather than general software defects.

\textit{Threats to Validity.} 
\toolname outperforms existing approaches on the dataset and real-world projects.
However, due to the reliance on GDB and LLDB and our implementation, \toolname currently supports only C programs, disabling the experimental comparison with tools designed for other languages like Java or Python.
With the help of equivalent tools in other languages, we plan to support more programming languages in the near future.

\section{Related Work}

In this section, we will introduce the relevant work on automated vulnerability repair and the code tasks based on the LLM agent. 

\subsection{Automated Vulnerability Repair}

Traditional vulnerability repair~\citep{gao2021beyond, huang2019using, song2024provenfix} efforts often rely on techniques such as symbolic execution~\citep{king1976symbolic} and program synthesis~\citep{jha2010oracle} to generate patches based on patterns.
While this approach has yielded some success, the patches produced often lack flexibility and accuracy, struggling to address the increasingly complex and varied types of vulnerabilities encountered today.
With the advancement of deep learning, some works based on Neural Machine Translation (NMT) have shown promising results~\citep{chi2022seqtrans, fu2022vulrepair, chen2022neural}. 
For example, \VulRepair~\citep{fu2022vulrepair} utilizes the CodeT5~\citep{wang2021codet5} framework, incorporating a Byte Pair Encoding (BPE)~\citep{sennrich2015neural} tokenizer.
However, these methods are limited by the capabilities of the models and the insufficiency of datasets, with accuracy rates below 25\%. 
Moreover, the actual repair capabilities of NMT-based methods are strongly tied to the datasets they were trained on, which further diminishes their effectiveness on untrained data.
Our approach leverages the understanding and generation capabilities of LLMs, offering higher accuracy and greater scalability compared to the aforementioned methods, capable of repairing various types of vulnerabilities.

\subsection{Code Tasks based on LLM Agent}

LLM agent~\citep{zhao2024expel} represents an innovative application of its capabilities, autonomously planning and executing actions to fulfill specific objectives. 
The fundamental mechanism involves providing the LLM with a prompt detailing the current state of the environment, the desired goal, and possible subsequent actions.
The model then determines the most appropriate action to take. 
To enhance the capabilities of LLMs in code-related tasks, existing research has already introduced various proven techniques.

\textbf{Chain-of-thought.} 
CoT~\citep{wei2022chain} has been proposed to improve the ability of LLMs to perform complex reasoning.
CoT enables the LLM to elicit multi-step reasoning behavior by decomposing multi-step problems into intermediate steps, which improves performance by a large margin on arithmetic reasoning. 
Previous research~\citep{yin2024thinkrepair, xia2023keep} has employed the CoT approach for program repair, providing an interpretable window into the behavior of LLMs.
This method effectively enhances the logic generated by LLMs.
However, it still heavily relies on the inherent capabilities of the LLMs and does not contribute additional knowledge to the repair process undertaken by LLMs.

\textbf{Retrieval-augmented generation.} 
Retrieval-augmented generation (RAG) is a model that combines information retrieval with text generation, and the Retrieval-Augmented Code Generation (RACG) technique enhances LLMs by retrieving code snippets or structures from code repositories~\citep{jiang2024survey}. Currently, RRAG technology has become a common technique for LLM agents tackling code-related tasks, spawning numerous research studies based on this approach~\citep{zhang2024llm, liu2024codexgraph, zhang2023repocoder, pan2024enhancing, arora2024masai, liu2024marscode}. For example, \autocoderover{}~\citep{zhang2024autocoderover} represents programs as abstract syntax trees to enhance LLMs' understanding of the root causes of issues.
The approach of using RRAG technology to retrieve static code information provides static insights for software engineering tasks.
However, this method alone is insufficient for effectively completing vulnerability repair tasks.
\MarsCodeAgent{}~\citep{liu2024marscode} exhibits good performance on the SWE-Bench~\citep{jimenez2023swe} by integrating dynamic debugging. 
However, its debugging effectively gathers test outcomes rather than capturing the runtime state of the program at the moment a vulnerability is triggered. 
Although these details are beneficial, remedying vulnerabilities necessitates more pivotal information. 
Our method acquires the program's runtime state using debugging tools and leverages the CFC to suggest expected states for comparison by the LLM. 
This approach not only gathers more critical information but also deepens the understanding of vulnerabilities, naturally leading to more effective repairs.

\section{Conclusion}

In this paper, we proposed \toolname, a novel dynamic state-aware agent for automated vulnerability repair.
It obtains the actual state of the program through program debugging and perceives the expected state based on the crash-free constraint.
Through the continuous comparison of these two states, \toolname attains a more profound understanding of the vulnerabilities, thereby facilitating the generation of more accurate and effective patches.
We selected 50 real-life projects with vulnerabilities, and \toolname successfully fixed \projectprecision of them, significantly outperforming existing approaches.

\bibliographystyle{ACM-Reference-Format}
\bibliography{reference}

\clearpage
\appendix

\section{Specific experimental data}

The detailed result of \toolname is given in Table~\ref{tab:detailed_results}.

\begin{table}[h]
    \footnotesize
    \vspace{-6pt}
    \centering
    \captionsetup{justification=centering}
    \captionsetup{skip=0pt} 
    \caption{Detailed results on each vulnerability \\ SE: Semantically Equivalent, P: Plausible, F: Failed}
    \vspace{-10pt}
    \begin{tabular}{ccccccc} \\
        \toprule
        Subject       & CVE/OSS-Fuzz ID   & \makecell{Root\\Cause} & \makecell{Fix\\Location} & \makecell{Patch\\Generation} &
        Tokens (Cost) & Time (s) \\
        \hline
        libtiff       & CVE-2016-3623     & \ding{51} & \ding{51} & SE  & 39.3K (\$0.120)     & 273   \\
        libtiff       & CVE-2016-5321     & \ding{51} & \ding{51} & SE  & 51.2K (\$0.160)     & 129   \\
        libtiff       & CVE-2016-10094    & \ding{55} & \ding{51} & P   & 55.5K (\$0.197)     & 183   \\
        libtiff       & CVE-bugzilla-2633 & \ding{55} & \ding{55} & P   & 46.3K (\$0.178)     & 303   \\
        libxml2       & CVE-2012-5134     & \ding{51} & \ding{51} & SE  & 25.8K (\$0.095)     & 81    \\
        libxml2       & CVE-2016-1838     & \ding{51} & \ding{51} & P   & 36.1K (\$0.113)     & 525   \\
        libxml2       & CVE-2016-1839     & \ding{51} & \ding{51} & SE  & 194K (\$0.556)      & 140   \\
        libxml2       & CVE-2017-5969     & \ding{51} & \ding{51} & P   & 46.5K (\$0.147)     & 109   \\
        libjpeg-turbo & CVE-2012-2806     & \ding{55} & \ding{55} & P   & 25.6K (\$0.086)     & 265   \\
        libjpeg-turbo & CVE-2017-15232    & \ding{51} & \ding{51} & SE  & 27.4K (\$0.103)     & 225   \\
        libjpeg-turbo & CVE-2018-14498    & \ding{55} & \ding{55} & F   & 117.2K (\$0.372)    & 398   \\
        libjpeg-turbo & CVE-2018-19664    & \ding{51} & \ding{55} & F   & 51.3K (\$0.192)     & 304   \\
        jasper        & CVE-2016-8691     & \ding{51} & \ding{51} & SE  & 815.1K (\$2.44)     & 383   \\
        jasper        & CVE-2016-9387     & \ding{51} & \ding{55} & F   & 71.5K (\$0.249)     & 300   \\
        elfutils      & 43307             & \ding{51} & \ding{51} & F   & 87.5K (\$0.231)     & 247   \\
        libplist      & 44393             & \ding{51} & \ding{51} & P   & 73.2K (\$0.193)     & 245   \\
        libplist      & 44574             & \ding{51} & \ding{51} & SE  & 23.1K (\$0.068)     & 167   \\
        elfutils      & 45628             & \ding{51} & \ding{51} & SE  & 115.0K (\$0.299)    & 322   \\
        file          & 47961             & \ding{51} & \ding{55} & SE  & 41.1K (\$0.108)     & 259   \\
        libcoap       & 48362             & \ding{51} & \ding{55} & F   & 34.4K (\$0.093)     & 234   \\
        file          & 48736             & \ding{51} & \ding{51} & P   & 98.9K (\$0.254)     & 307   \\
        file          & 51608             & \ding{51} & \ding{51} & SE  & 8.5K (\$0.025)      & 215   \\
        ndpi          & 52229             & \ding{55} & \ding{51} & SE  & 197.7K (\$0.501)    & 678   \\
        cups          & 54069             & \ding{51} & \ding{51} & F   & 133.8K (\$0.356)    & 357   \\
        libplist      & 54948             & \ding{51} & \ding{51} & F   & 113.5K (\$0.294)    & 373   \\
        libplist      & 55035             & \ding{51} & \ding{51} & F   & 43.1K (\$0.120)     & 257   \\
        libjpeg-turbo & 55413             & \ding{51} & \ding{51} & P   & 75.9K (\$0.205)     & 262   \\
        libxml2       & 55980             & \ding{51} & \ding{51} & F   & 392.5K (\$0.994)    & 270   \\
        libxml2       & 57410             & \ding{51} & \ding{51} & F   & 108.4K (\$0.287)    & 328   \\
        cyclonedds    & 57614             & \ding{55} & \ding{55} & F   & 73.6K (\$0.187)     & 134   \\
        ndpi          & 59393             & \ding{51} & \ding{51} & SE  & 21.9K (\$0.064)     & 208   \\
        file          & 59438             & \ding{51} & \ding{51} & SE  & 53.5K (\$0.149)     & 275   \\
        libxml2       & 61337             & \ding{51} & \ding{51} & SE  & 54.7K (\$0.146)     & 299   \\
        libxml2       & 62886             & \ding{55} & \ding{55} & F   & 301.5K (\$0.778)    & 465   \\
        lcms          & 63954             & \ding{51} & \ding{51} & SE  & 34.6K (\$0.095)     & 166   \\
        libxml2       & 65120             & \ding{55} & \ding{55} & F   & 12.8K (\$0.034)     & 1584  \\
        gpac          & 65209             & \ding{55} & \ding{55} & F   & 43.6K (\$0.111)     & 159   \\
        gpac          & 65215             & \ding{51} & \ding{51} & SE  & 7.6K (\$0.022)      & 293   \\
        ndpi          & 65362             & \ding{51} & \ding{51} & F   & 224.3K (\$0.576)    & 744   \\
        htslib        & 65383             & \ding{55} & \ding{55} & F   & 1520.6K (\$3.816)   & 1120  \\
        gpac          & 66032             & \ding{55} & \ding{55} & F   & 79.7K (\$0.209)     & 595   \\
        gpac          & 66187             & \ding{51} & \ding{51} & P   & 22.8K (\$0.065)     & 368   \\
        gpac          & 66196             & \ding{51} & \ding{51} & SE  & 101.8K (\$0.266)    & 392   \\
        htslib        & 66369             & \ding{55} & \ding{55} & F   & 125.9K (\$0.318)    & 142   \\
        gpac          & 66415             & \ding{51} & \ding{51} & SE  & 46.5K (\$0.122)     & 248   \\
        gpac          & 66591             & \ding{55} & \ding{55} & F   & 32.6K (\$0.083)     & 355   \\
        gpac          & 66696             & \ding{51} & \ding{51} & P   & 73.6K (\$0.189)     & 274   \\
        gpac          & 66742             & \ding{51} & \ding{51} & P   & 20.4K (\$0.055)     & 294   \\
        gpac          & 67354             & \ding{51} & \ding{51} & SE  & 20.5K (\$0.055)     & 267   \\
        ndpi          & 67881             & \ding{51} & \ding{51} & F   & 25.1K (\$0.070)     & 222   \\
        \bottomrule
    \end{tabular}
    \vspace{-20pt}
    \label{tab:detailed_results}
\end{table}

\end{document}